# Temperature Dependent Heterogeneous Rotational Correlation in Lipids

*Neda Dadashvand [a] and Christina M. Othon [a, b]*

[a] Department of Physics, Wesleyan University, Middletown CT 06457 USA

[b] Molecular Biophysics Program, Wesleyan University, Middletown CT 06457 USA

**Graphical Abstract:**

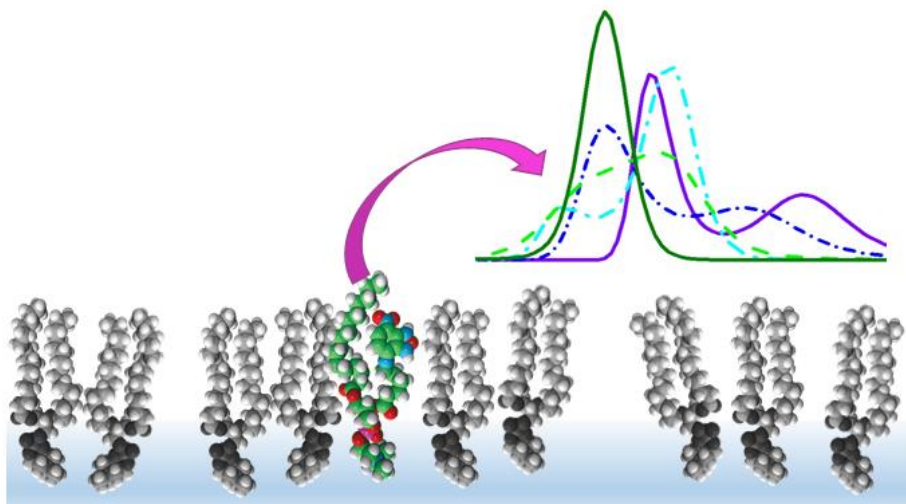

author email: ndadashvand@wesleyan.edu , cothon@weseleyan.edu

Corresponding author: cothon@wesleyan.edu

**Abstract**

Lipid structures exhibit complex and highly dynamic lateral structure; and changes in lipid density and fluidity are believed to play an essential role in membrane targeting and function. The dynamic structure of liquids on the molecular scale can exhibit complex transient density fluctuations. Here the lateral heterogeneity of lipid dynamics is explored in free standing lipid monolayers. As the temperature is lowered the probes exhibit increasingly broad and heterogeneous rotational correlation. This increase in heterogeneity appears to exhibit a critical onset, similar to those observed for glass forming fluids. We explore heterogeneous relaxation in in a single constituent lipid monolayer of 1,2-dimyristoyl-sn-glycero-3-phosphocholine (DMPC) by measuring the rotational diffusion of a fluorescent probe, 1-palmitoyl-2-[1]-sn-glycero-3-phosphocholine (NBD-PC), using wide-field time-resolved fluorescence anisotropy (TRFA). The observed relaxation exhibits a narrow, liquid-like distribution at high temperatures ($\tau \sim 2.4$ ns), consistent with previous experimental measures [1, 2]. However, as the temperature is quenched, the distribution broadens, and we observe the appearance of a long relaxation population ($\tau \sim 16.5$ ns). This supports the heterogeneity observed for lipids at high packing densities, and demonstrates that the nanoscale diffusion and reorganization in lipid structures can be significantly complex, even in the simplest amorphous architectures. Dynamical heterogeneity of this form can have a significant impact on the organization, permeability and energetics of lipid membrane structures.

The complex, heterogeneous, and dynamic structure of natural lipid membranes appears to play a critical role in essential functional inter/intracellular processes. Very recently it has become apparent that even simple single constituent and binary mixture lipid structures exhibit complex transient nanoscale lateral heterogeneity [1, 3-10]. Experimental demonstrations include: translational correlation lengths with size of ~30 Å and lifetimes of tens of nanoseconds as measured by neutron backscattering [5, 8]; the heterogeneous lateral diffusion of fluorescent probes in binary lipid compositions [11-13], the broad distribution of steady-state single molecule anisotropy in the liquid-ordered state of raft-forming compositions [14]; and polymer lateral diffusion in a supported lipid bilayer in the disordered phase [15]. Previously we demonstrated a direct relationship between the degree of lateral heterogeneity of the rotational correlation of a fluorescence probe and the lateral packing density of lipids for the liquid phase of DMPC [1]. A shoulder appeared in the dynamic distributions for the highest molecular packing fractions. This form of dynamical heterogeneity bears striking resemblance to the lateral heterogeneity observed for colloidal systems at high packing fraction [16, 17]. Here we explore the dynamical heterogeneity as the temperature is quenched and approaching the phase transition for moderate surface tensions. This exploration of the phase space demonstrates the appearance of a population of probes which experience a high density local environment, though on the macroscale the material is amorphous and no heterogeneity is detectable.

Heterogeneous liquid dynamics are most obvious at low temperatures as the glass transition is approached due to the extension of clustered state lifetimes, however non-exponential diffusion and cooperative motion has been observed for many liquids far from the glass transition [18-20]. Dynamic heterogeneity observed in fluids is occasionally referred to as a dynamic glass transition. The transition is not abrupt, instead exhibiting a continuous slowing down of motion corresponding to a growth in correlation size as the glassy or gel transition is approached. The most common signature of complex dynamics is a stretching of the dynamic relaxation over large timescales indicating heterogeneity of lateral density throughout the fluid. The lifetime of lateral density fluctuations has been modelled using coarse-grained simulation in single constituent lipid bilayers [9, 10, 21]. The dynamic fluctuations are very small (~ 10 nm) and short lived (10 μs).

The extreme size and brief lifetime of such structures dramatically limits the ability to directly detect the presence of lateral structural heterogeneity. In fact, most single molecule measures of lateral diffusion have a time resolution ~10 ms, which averages over any transient lateral heterogeneity only allowing measures of average lateral diffusion rates for small molecules such as lipid probes or nanoparticles. To overcome this obstacle we employed wide-field Time-Resolved Fluorescence Anisotropy (TRFA). The time resolution of our system is ~40 ps, providing excellent sensitivity to changes in the rotational diffusion of lipids which take place on the order of a few ns in most systems [2, 22-24]. This reorientation is over 2 orders of magnitude faster than the anticipated lifetime of density fluctuations in high density lipid states. By measuring the distribution of probe molecule rotational correlation times we provide a measure of the dynamical heterogeneity within this densely packed amorphous fluid. Condensed matter theory predicts that such collective motion should depend critically on both the free volume of the liquid and the temperature of the system [16, 19, 25], and thus the distribution of rotational correlation should become more heterogeneous as the temperature of the system is reduced.

We chose to make measurements on a free-standing lipid monolayer, on which we can directly control the lateral density by compressing the film to small surface areas. The absence of a support structure eliminates the interfacial interactions as the source of lateral inhomogeneity. We make use of a monolayer of lipids which assemble at the air-water interface. The Langmuir technique has proven extraordinarily useful in monitoring and detecting phase transitions of lipids at the air-water interface. For the purpose of this study we wish to monitor changes in the diffusion of molecules in the amorphous phase as they approach the transition to the condensed state. Using both optical microscopy and a Wilhelmy film balance; we can directly monitor our proximity to the phase transition and directly detect any change in probe partitioning or phase separation. Using a solvochromatic fluorescent probe which exhibits a strong emission shift with respect to changes in solvent polarity [2, 26, 27] and water accessibility ensures that the probe dynamics are due to changes in local viscosity, and not due to changes in probe orientation or location within the lipid structure.

To assess the temperature dependent heterogeneity of rotational diffusion, we must first determine a span of phase space that allows us to examine the largest range of temperature change without undergoing a large-scale probe phase separation. We could consider two control parameters, either constant area per molecule or constant surface tension. A constant area model has the disadvantage that local clustering or changes in molecular ordering could transpire on the surface during the measurement, without an active direct measurement we would be unaware of the change in organization. The maintenance and feedback used to maintain constant surface tension is easily implemented to directly measure and actively control. This approach allows us to independently measure any loss of area or hysteresis as an indication of a significant change in molecular ordering or clustering.

Our previous results indicated that for samples at room temperature, lipid monolayers began exhibiting a broad distribution of dynamics at an area per/lipid below 75 Å$^2$/molecule [1]. This corresponds to a surface tension of approximately 15 mN/m near 30 °C (Figure 1). The presence of our fluorescent probe appears to slightly lower the critical temperature of the system; this is consistent with the impact of impurities within lipid phases. We chose to measure dynamics at a surface tension of 15 mN/m for temperature temperatures between 9.5 °C to 30 °C. This temperature range is bound by the ability of our water bath to cool our Langmuir trough, and at the upper end by solvent evaporation and thermal effects during TRFA acquisition. At pressure 15 mN/m we anticipated that for the lowest temperature the DMPC monolayer is just entering the coexistence region of the LE-LC phase while for higher temperatures the layer is in a single liquid phase. We observe a broad coexistence regime for DMPC over molecular areas from 50-60 Å$^2$/molecule, consistent with previous observations [28].

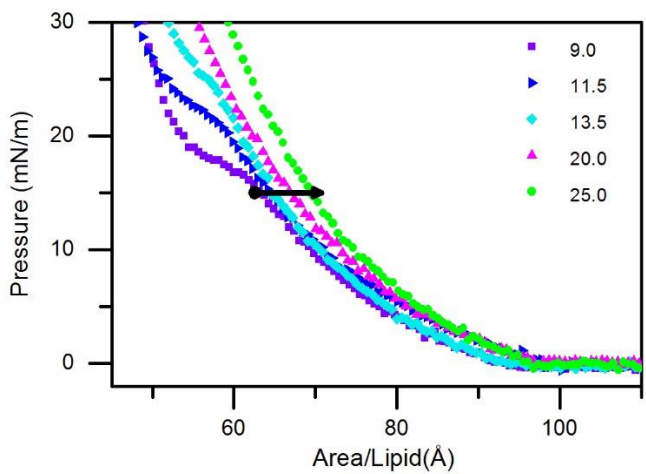

*Figure 1 Pressure-area isotherms of DMPC labelled with 1 mol% fluorescent probe. Temperature increases from 9.5 °C to 30 °C from left to right. The arrow represents the region of phase space explored in this study for temperature dependent dynamical heterogeneity.*

To verify that the probe remains homogenously distributed throughout the liquid phase we used optical microscopy. The plateau in the isotherm of DMPC (Figure 1) is an indication of phase coexistence of the expanded liquid phase and the condensed phase of DMPC. Fluorescence images were taken at all temperatures for which the rotational diffusion was measured, see Figure 2. For temperatures above 15°C the image was featureless, exhibiting a single liquid phase over the entire monolayer surface. For 9.5°C we observed phase coexistence at a surface tension of 15 mN/m. This result was initially surprising, as the isotherms indicated that at this temperature and pressure the system would be just above the onset of phase separation. However, it has been shown that the critical pressure for the phase transition can be rate dependent [29, 30]. The isotherms were measured at a compression rate of 5.3 cm2/s. However, the images at 9.5 °C was acquired after sitting at constant pressure for 15 minutes in a manner consistent with our acquisition of TRFA signal. The domains began to form after a small amount of time indicating a slow, kinetically limited phase separation. The imaging of layers at 11.5 °C and 13.5 °C exhibited phase separation after being held at 15 mN/m for an extended period of time (>30 minutes). No phase separation was observed for temperatures above 15°C, which were observed over extended time periods (>2 hours). This imaging coupled with time-resolved and steady state measures of the fluorescent probe properties, ensures that the probe is partitioned only to the liquid expanded phase throughout the study; only at the lowest temperatures are there any indication of a lateral phase separation.

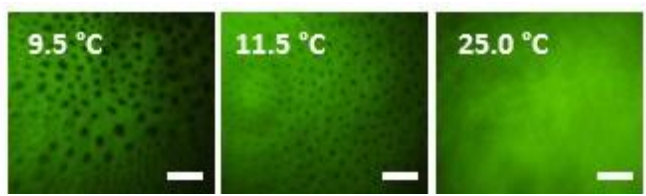

*Figure 2 Fluorescence microscopy images of DMPC monolayers labelled with 1mol% NBD-PC for various temperatures. Images were taken after 15 minutes of being held at constant surface pressure of 15 mN/m. Scale bar represents 50 μm.*

We have measured the steady state fluorescence spectrum of DMPC monolayers labelled with 1 mol% NBD-PC at surface pressures across the entire range of pressures explored. We measure the fluorescence in-situ using an Ocean Optics QE65000-FL Fluorescence Spectrometer. Fluorescence was collected through the Cassegrain objective, then focused onto a 600 μm diameter fiber optic which is connected directly to the spectrometer. Fluorescence is collected for 2 seconds and the data represent the average of 5 scans. Emission is detected at 543 nm for all the data sets, consistent with the emission maximum observed for NBD-PC in the literature [27]. This consistency in fluorescence maximum is one indication that the probe maintains a well anchored position through the compression range observed. If there were a large change in location of the probe within the layer it would experience a large shift in solvent polarity and solvent accessibility of the probe which would result in a significant spectral shift [27].

The time-resolved fluorescence anisotropy and lifetime of the probe were measured on our wide-field time-resolved fluorescence anisotropy microscope as described elsewhere [1]. The excitation is tuned to 463 nm by frequency doubling (LBO Type I, 5×5×2mm) the output of a tunable ultrafast oscillator (Coherent Chameleon Ultra II, 140 fs). The pulse repetition rate is reduced using a pulse selector (Conoptics model 305). The final excitation beam has a power of 20 mW at 10 MHz. We employ a wide-field excitation at the air-water interface with a beam diameter of ~500 μm. The fluorescence signal is collected through an all reflective, Cassegrain objective (Newport model 50105-02, 15x, Infinite BFL), the signal is separated into the two polarization orientations using a polarizing beam splitting cube, and focused through the monchromators and onto the single photon avalanche photodiodes (SPAD, id100 ID Quantique). The signal is passed to a router (HRT-41) and TCSPC module (SPC-1300 Becker-Hickl), and the rotational correlation is measured for DMPC monolayers labelled with 1 mol% NBD-PC for various temperatures at a surface pressure of 15 mN/m.

The TRFA signal is analysed by calculating the rotational anisotropy. The difference between the parallel polarization channel and the orthogonal polarization channel represents a measure of the ensemble reorientation of the probe molecules as a function of time. The well anchored position of NBD within the head group region of the layer and the well-defined excitation/emission dipole moment of the probe provides a reliable and repeatable rotational correlation measure. This probe provides a size similar to the DMPC molecule itself, and its rotational correlation is dominated by the density and viscosity of the local environment. The rotational anisotropy decays are analyzed using a Maximum Entropy Method (MEM) analysis [31, 32]. This approach makes no assumptions about the relaxation modes present in the data, presenting an unbiased analysis of the rotational correlation distributions.

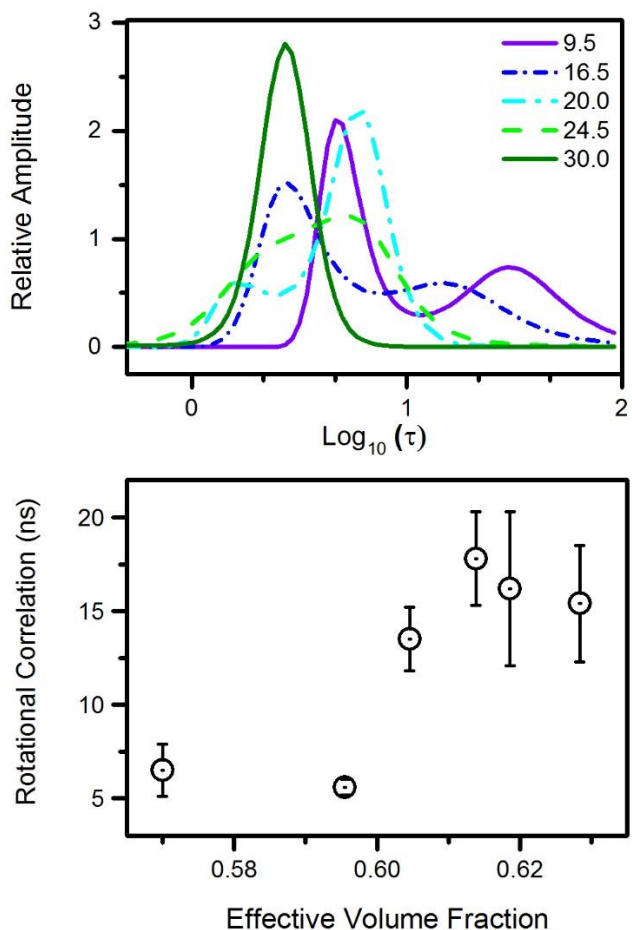

*Figure 3 (Top Panel) Probe rotational correlation distributions of labelled DMPC monolayers held a constant surface tension as temperature is varied from 9.5 – 30 °C. The distributions become increasingly heterogeneous as the phase transition is approached. (Lower Panel) Rotational correlation times of the slowest population increases with effective volume fraction of the lipid.*

At the highest temperatures we measure a very narrow, exponential relaxation of the probe which indicates a single low density liquid-like diffusion with a time constant of 2.62+/-0.4 ns at 30 °C. As the temperature is lowered the rotational correlation time extends and broadens, see Figure 3. This broadening is indicative of a heterogeneity in the microenvironments sampled by the probe.

This broadening of the dynamic distribution at high packing fraction is a signature of heterogeneity as would be expected for highly correlated motion. As we further quench the temperature towards the phase transition at this surface tension, the broad distribution splits into a bimodal distribution. This bimodal distribution remains unaltered throughout the phase coexistence regime (Figure 4), which is consistent with the behavior observed previously for the phase coexistence regime of DPPC [1]. As the surface tension pushes the monolayer through the phase coexistence regime, the probe remains partitioned solely to the liquid phase of the lipids. The average area/molecule of this phase does not change through the phase coexistence regime, and the measured dynamics remain consistent throughout. The time constants for our bimodal distribution in the phase coexistence regime also remain consistent as the temperature is lowered. The observance of a bimodal relaxation was not anticipated. It appears to be a robust feature of the lipid collective dynamics. As the average area per lipid is reduced either by increasing the lateral density monolayer [1] or by reducing the temperature and allowing the layer to condense at constant surface tension; the dynamics appear to become increasingly heterogeneous, eventually displaying a bimodal distribution at the highest packing fractions.

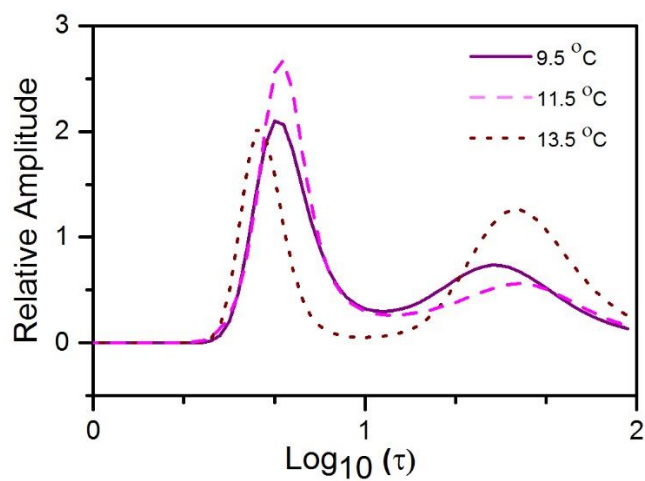

*Figure 4 Distribution of probe rotational correlation times for the low temperature phase coexistence regime.*

An abrupt onset of heterogeneous dynamics is observed for the time constant scaling with packing fraction. To make this estimation we used the average area per lipid of crystalline DMPC as the limiting packing area for lipid molecule [33]. We plot the time-scale for the slowest population dynamics as a function of the packing fraction in Figure 3 lower panel. We observe a critical packing fraction of ~0.59 for the appearance of the high-density, low mobility state. Below this density the dynamics exhibit a single Arrhenius-like diffusion distribution.

Biological membranes segregate into specialized functional domains of distinct composition, resulting in measurements of diffusion constants that differ by a factor of ten, depending on the length scale probed [34]. Dynamic clustering of the type demonstrated here would dramatically impact the dynamics of particles with a biological membrane and would affect membrane recognition, permeability, transport, and protein self-assembly. The densities explored here are below those found in natural membranes, and therefore such density fluctuations could play a significant role in the dynamic function of these structures. The presence of molecules of slightly different size and shape with different concentration will alter the dynamic distribution and enable tuning of local density and dynamic heterogeneity within lipid structures. The transient dynamic structure of amorphous liquids can be complex and should be taken in to consideration when modelling energetic interactions within crowded lipid structures.

**Acknowledgments**

This work was partially supported by Wesleyan Grants in support of scholarship. We would like to thank F. Starr, D. Oliver, and E. Vega-Lozada for discussions.